# Eavesdropping risk evaluation for non-line-of-sight terahertz channels by metallic wavy surface in rain

**Peian Li,[1] Wenbo Liu,[1] Da Li,[1] Mingxia Zhang,[1] Xiaopeng Wang,[3] Houjun Sun,[1,2] and Jianjun Ma[1,2,*]**

[1]*Beijing Institute of Technology, Beijing, 100081, China*
[2]*Beijing Key Laboratory of Millimeter and Terahertz Wave Technology, Beijing Institute of Technology, Beijing, 100081, China*
[3]*Radar Operation Control Department, Beijing, 100044, China*

**Abstract:** Non-line-of-sight (NLOS) data transmission through surface reflection is pivotal for enhancing the reach and efficiency of terahertz (THz) communication systems. However, this innovation also introduces significant eavesdropping risks, exacerbated by the complex bistatic scattering effects during adverse weather conditions like rain. This work delves into the assessment of the vulnerabilities of NLOS THz communication channels to eavesdropping under simulated rain conditions using metallic wavy surfaces (MWS). The observation reveals the feasibility of successful signal interception under these conditions, highlighting a prevalent security concern for outdoor terahertz communication networks utilizing NLOS channels to broaden coverage. This insight underscores the critical need for addressing and mitigating potential eavesdropping threats to ensure secure and reliable terahertz communications in varied environmental conditions.

## 1. Introduction

At the pioneer of wireless communication technology, terahertz (THz) frequencies (100 GHz to 10 THz) stand as a revolutionary innovation, signaling the arrival of an era characterized by unparalleled data transmission speeds and minimal delays. These frequencies are at the heart of revolutionizing future wireless networks, embodying the vision of seamless connectivity [1]. Unlike sub-6 GHz signals, which have longer wavelengths and can diffract around obstacles and penetrate various materials, THz waves' shorter wavelengths result in more line-of-sight (LOS) propagation and limited diffraction. Consequently, THz signals more confined to directed paths. This directionality reduces the risk of interception by unintended receivers, as eavesdropping would require precise alignment with the signal path [2, 3]. Yet, despite this benefit, signal privacy is still one of the major concerns, especially with the potential of channel scattering to undermine secure transmission [4].

As the domain of wireless communication continues to evolve, the imperative for robust security strategies to protect data from unauthorized interception becomes increasingly pronounced. Recent progress has shed light on specific susceptibilities tied to THz frequencies, prompting an urgent reassessment of eavesdropping risks [5]. Intriguingly, it has been disclosed that, notwithstanding the intrinsic security attributes of THz frequencies, refined approaches - such as positioning cylindrical reflectors inside channel path - can stealthily enable eavesdropping [6]. This threat intensifies in scenarios where adversaries leverage the spatial dynamics of wireless channels, intensifying the challenge of safeguarding data privacy [7-10]. To counteract this, the adoption of metallic wavy surfaces (MWS) to modify THz channel path has been proposed as an innovative countermeasure, even though introducing other eavesdropping susceptibilities [11].

Investigations into scattering-based non-line-of-sight (NLOS) data transmission have significantly advanced propagation performance [12]. The deployment of intelligent reflecting surfaces (IRS), for instance, facilitates signal direction control via phase shift adjustments, managed by a pre-set controller. This approach augments both the security and efficiency of signal transmission [13, 14]. Moreover, the adaptability of unmanned aerial vehicles (UAV) equip them with a pivotal role in relay-based wireless systems, boosting coverage, dependability, energy efficiency, and network bandwidth [15]. However, the efficacy of these technologies often hinges on the precise performance of surfaces [16]. Environmental influences such as surface contamination, oxidation, mechanical vibrations, and variations in temperature and roughness can alter surface properties, endangering channel propagation [17, 18]. Traditional discussions on THz channel transmission have tended to overlook the effects of environmental conditions, assuming ideal scenarios. Nonetheless, adverse weather phenomena like fog, snow, and rain, significantly impair THz wireless channels, inducing substantial signal attenuation and scattering [19-23].

This problem is worsened by the build-up of atmospheric particles on critical signal-reflecting surfaces. Although theoretical frameworks have been established to evaluate the security performance of such conditions for line-of-sight THz channels [24, 25], a noticeable absence in empirical research persists, especially concerning the efficacy of NLOS outdoor THz channels in such environments [26].

This work endeavors to bridge the research gap by specifically investigating the eavesdropping risks associated with NLOS THz channels that utilize MWS during rainfall. These surfaces are distinct from IRS, as they do not alter signal phase or intensity, serving solely to modify signal paths through their physical structure. Recognizing the challenges posed by fluctuating and uncontrollable outdoor weather conditions, we employ a controlled laboratory environment to emulate rain, allowing for precise manipulation and measurement of environmental variables. This approach enables a detailed examination of how rainfall affects signal scattering and absorption, which are critical to understanding the vulnerability of these channels to eavesdropping. The primary contribution includes: 1) quantitative assessments of how different surface conditions (smooth vs. wavy) under rainfall influence the susceptibility of THz communications to eavesdropping. This includes measuring signal scattering, path loss, and potential interception points; 2) exploring an under-researched area of THz channel behavior, offering new insights into surface-engineered security enhancements for NLOS communications; 3) revealing how specific configurations of surface geometry and environmental conditions (rainfall intensity and drop size distribution) affect the integrity and security of transmitted channels. This work also contributes to the development of more secure THz communication strategies by elucidating the complex interplay between channel characteristics and environmental influences.

## 2. Experimental setup

The experimental design aimed to explore the impact of emulated rain on THz channel transmission and its susceptibility to eavesdropping. This setup included a THz transceiver system, a custom-designed MWS, and an advanced rain emulation module, as depicted in Fig. 1(a). The signal transmission component was equipped with a signal generator (Ceyear 1465D), a frequency multiplier (Ceyear 82406B), and a horn antenna (HD-1400SGAH25) featuring a dielectric lens with a 10 cm focal length. The receiving end mirrored this configuration, comprising an identical horn antenna and a power sensor (Ceyear 71718), mounted on an electronically controlled turntable for detecting scattering phenomena. This setup provides a frequency range of 110-170 GHz. Our measurements focus on the 140 GHz and 170 GHz continuous wave signal, with an emission power of 0 dBm. A key element of our setup was the MWS, connecting the transceiver to form NLOS channels, positioned 37 cm away from the transmitter (Tx) and 31 cm away from the receiver (Rx). These surfaces, fabricated from aluminum plates, were selected for their minimal surface roughness (~20 um), effectively minimizing roughness losses at frequencies below 1 THz. This was critical as aluminum, with its innate reflective properties, served as an almost ideal mirror in the THz spectrum without the need for additional polishing, which could minimize reflection losses that are typically more pronounced with common wall materials [11]. The surface featured one-dimensional sinusoidal ripples, with MWS (S2), illustrated in Fig. 1(a)'s inset, boasting an amplitude ($A$) of 0.7 mm and a period ($\varLambda$) of 6 mm. Aluminum's resilience to temperature shifts [16, 27, 28], as shown in Fig. 1(b), ensured its reflectivity remained consistent over a wide temperature range (-40 °C to 40 °C), validating its application across diverse outdoor scenarios. It is important to note that these MWS do not possess intelligent capabilities, such as phase or intensity modulation of signals, which are characteristic of IRS. Our use of MWS focuses on their geometric and reflective properties to understand how they influence signal scattering and security in THz communications.

At the heart of our rain emulation was a polyethylene (PE) rain generator, positioned 105 cm above the setup. This device, with a dimension of 17 cm × 11 cm × 5 cm (length × width × height), was designed with a water inlet and 98 outlets for precise rain emulation. Its location could be adjusted to vary the rain-affected (raining) area, facilitating studies on direct surface wetting and path interception. Accompanied by a pressure pump and flow meter, this system allowed for precise control over rainfall intensity, anchored securely to ensure both flexibility and stability during experiments. To ensure clarity in our findings, we utilized an exaggerated rainfall rate of 8065 mm/hr, a necessity given the experimental constraints in channel length and aimed at distinguishing the effects of droplet accumulation versus channel interception. Observations in Fig. 1(a) revealed that the smooth surface (S1) collected raindrops in a disordered manner, whereas the MWS (S2) encouraged a uniform distribution, attributable to the altered surface tension and contact angles induced by its structure, forming a cohesive water film [29]. This resembles the hydrophobicity of surfaces with grooves, where the anisotropic geometry allows the liquid drop to preferentially spread along the groove [30].

The eavesdropping assessment configuration, shown in Fig. 1(c), facilitated communication between Alice (the Tx) and Bob (the legitimate Rx) across the MWS, positioned to emulate real-world interactions. Both the MWS and receiver were mounted on rotating platforms, enabling precise adjustments to channel angles and the collection of scattering data with an angular step of 1°. The experimental design included two distinct raining scenarios: one affecting the surface alone (Rn1 in Fig. 1(c)) and the other impacting the channel path between Tx and the Sx (Rn2 in Fig. 1(c)). Conducted under rigorously controlled conditions, this setup minimized external variable interference, ensuring the reliability and accuracy of our eavesdropping vulnerability assessment. Here, we positioned both the Tx and Rx outside the raining area, focusing specifically on the influence of rain on the channel itself rather than on the transceiver devices. This approach was chosen to simulate conditions where, in practical scenarios, the critical components such as Tx and Rx are often protected from direct exposure to environmental elements, including rain, using shelters or other protective measures.

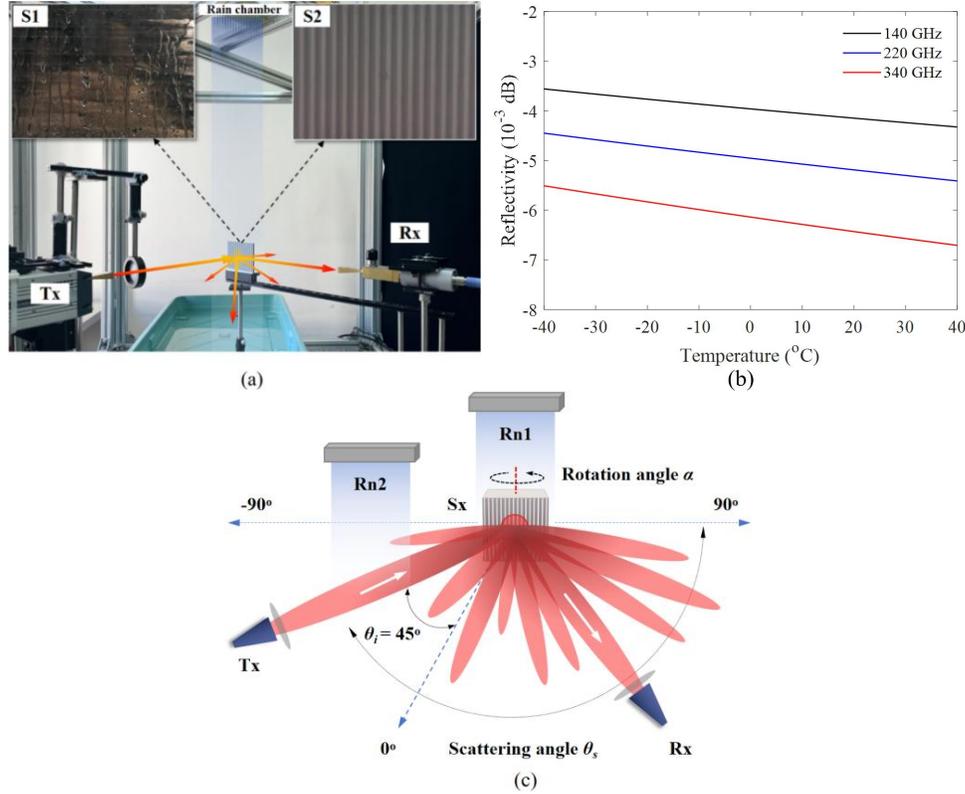

Fig. 1. (a) Picture of the eavesdropping measurement setup in rain. Inset: smooth surface (S1) and MWS (S2) with raindrops attached. The dashed lines are the guide lines, and the solid lines represent the signals. (b) Variation of reflectivity due to a flat aluminum plate with respect to temperature. (c) Schematic of the eavesdropping measurement in different raining area.

## 3. Channel performance analysis

In our investigation into the THz channel behavior under various weather conditions, we analyzed the impact of rainfall on signal reflection and scattering. Holding the incidence angle constant at 45º, we rotated the receiver to capture the scattering patterns under both clear (no rain) and rainfall conditions. Our analysis focused on comparing these patterns using two distinct surfaces: a smooth one, S1, and a MWS counterpart, S2, across three different transmission scenarios at 140 GHz, the lower boundary of THz frequency range [6]. It is worth noting that, in addition to the direction of specular reflection, the signal radiation from surface S2 exhibits multiple scattered signal peaks. These scattered signals are influenced by the surface structure and follow Bragg's diffraction law [11]. The conditions were labeled 'NoRain' for experiments without artificial rain generation, 'TxRain' for rain affecting the path between transmitter and surface, and 'SxRain' for rain directly impacting the surface alone, as described in Fig. 1(c). The results, shown in Fig. 2(a) and (b), revealed that rainfall notably increases scattering effects, a phenomenon that becomes significantly pronounced when rain directly impacts the surfaces ('SxRain' condition). The scattering, attributed to water droplets adhering to the surfaces [31], was markedly severe with the smooth surface S1, exhibiting much greater fluctuations than the noise level (-38 dBm). This suggests that the MWS S2 can be a superior option for NLOS channel reflection during rainfall.

To discern whether the observed scattering was a consequence of channel propagation through rain or the adherence of raindrops to the surfaces, we conducted further tests. These involved analyzing reflections and scatterings from both surfaces, with and without raindrop attachment, facilitated by a water spraying device to emulate consistent wetness. Fig. 2(c) clearly demonstrates that reflections off surface S1 were drastically altered by raindrop attachment, showing signal variations ranging from -3.94 dB to 13.18 dB. In contrast, the MWS S2 exhibited minimal impact, with most variations within a narrow range of -0.6 dB to 0.6 dB, highlighting the efficacy of the wavy design in mitigating water droplet adherence and its potential for outdoor deployment. In other words, the MWS can offer a predictable and stable performance across different environmental conditions. Such a consistent performance of MWS supports more reliable security measures in THz communications by reducing the variability in signal behavior that could otherwise be exploited by eavesdroppers.

Further examination into the scattering effects revealed that they originated both from droplets on the surface and those within the transmission path. Specifically, the scattering from raindrops in the air has been previously noted for its implications on eavesdropping risk [24]. Removing the surface to focus on the LOS channel affected by rainfall, we observed an amplification in signal scattering within a scattering range of 18° to 60° as shown in Fig. 2(d), peaking with a 7.1 dB increase, whereas the primary transmission area saw small power reduction, just around 1 dB. Returning to the 'TxRain' condition in Fig. 2(b), it became evident that the scattered multipath components from airborne raindrops could interact with the MWS under varying incident angles without causing discernible interference, aside from slight power reduction around 1 dB again. This observation led us to conclude that neither raindrops traversing the transmission path nor those adhering to the MWS critically undermined channel performance, affirming the stability of multipath transmission under these conditions.

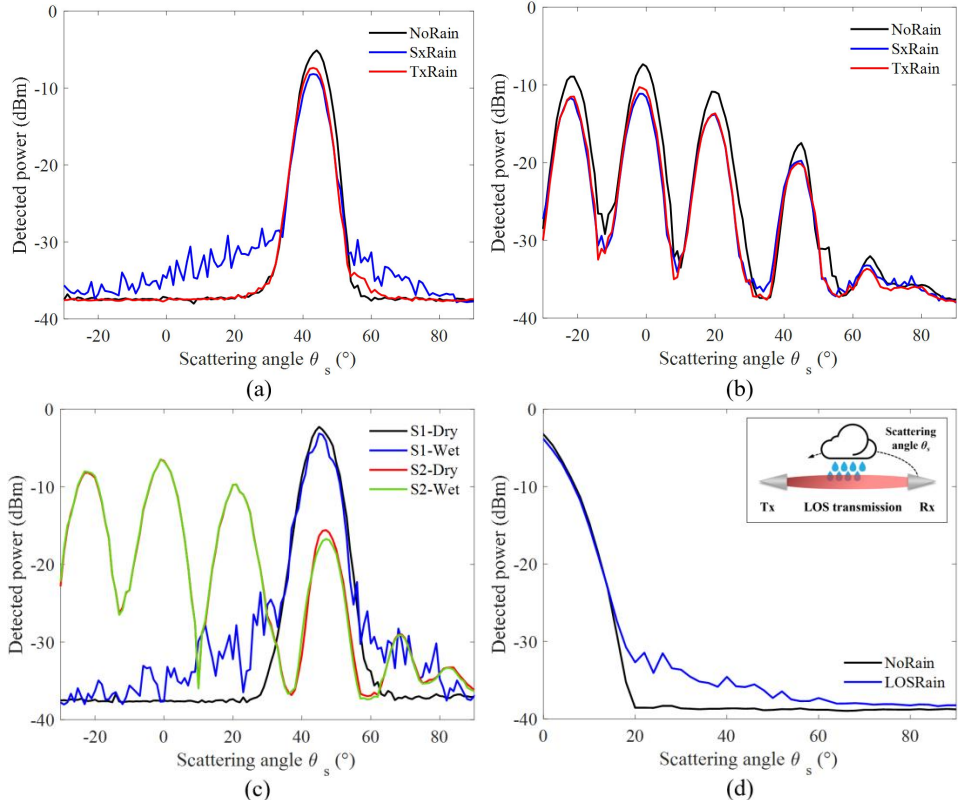

Fig. 2. (a) Measured scattering pattern after reflection by surface S1 under three conditions. (b) Measured scattering pattern after reflection by surface S2 under three conditions. (c) Angular distribution of the signal power for surface S1/S2 under dry and wet (attached rain droplets) conditions. (d) Angular distribution of power scattering from a line-of-sight channel without surfaces inserted. Inset: diagram of LOS transmission link under rainfall condition.

## 4. Eavesdropping performance analysis

In our analysis of eavesdropping potential under the setup described, we find that an eavesdropper, referred to as Eve, is capable of intercepting signals from various directions, including under conditions of rainfall. To quantitatively assess the risk and effectiveness of eavesdropping, we employed the concept of *normalized secrecy capacity* [6], a metric that provides insight into the absolute power levels of the transmitted signal. Alice, the transmitter, emits the signal at a 45° incidence angle ($\theta_i$) with a beam spot size of 3 cm, while Bob, the intended legitimate receiver, is situated to receive the specular reflection directly, also at a 45° angle ($\theta_s$). Eve attempts to intercept the transmission, rotating at one-degree intervals to measure and calculate the *normalized secrecy capacity* as per the given equation:

$$c_s = \frac{\log(1+SNR_{Bob}) - \log(1+SNR_{Eve})}{\log(1+SNR_{Bob})} \quad (1)$$

where SNR denotes the signal-to-noise ratio on a linear scale. A *normalized secrecy capacity* ($c_s$) value of one implies that Eve is unable to detect the signal, a value of zero indicates that Eve receives a signal strength equivalent to Bob's, and a negative value means that Eve's received signal strength surpasses that of Bob. For the purposes of our discussion, we establish a $c_s$ threshold of 0.5, adjustable based on Eve's decoding capabilities, under the assumption that eavesdropping becomes viable when $c_s$ falls below this value. Fig. 3(a) illustrates the angular positions where the *normalized secrecy capacity*, $c_s$, falls beneath the 0.5 threshold. These instances, occurring at each scattering peak, signify that the signals at these points, conforming to *Bragg's Law*, have adequate strength for interception and communication. This analysis not only underscores the potential for eavesdropping across a spectrum of angles but also highlights the criticality of signal security measures in mitigating unauthorized access, even in adverse weather conditions.

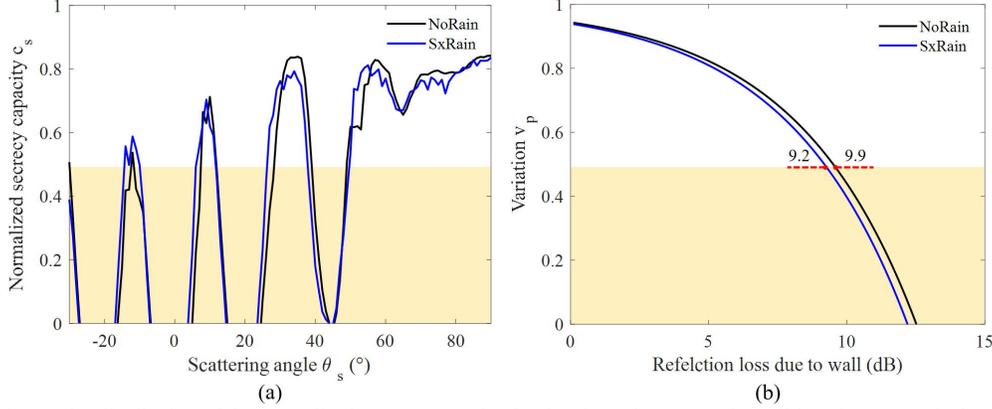

Fig. 3. (a) Angular distribution of the normalized secrecy capacity for the channels at several conditions is measured after scattering by the MWS S2. (b) Values of the variation computed for the values of reflection loss by typical surfaces for non-line-of-sight wireless connection under rainy and clear weather conditions. Yellow region means corresponding values below 0.5.

We introduce another crucial metric, termed *variation*, $v_p$, to quantitatively assess the change in signal power received by Bob due to the incorporation of a MWS in the NLOS channel.

$$v_p = 1 - \frac{SNR_{Bob}^{surface}}{SNR_{Bob}^{nosurface}} \quad (2)$$

The parameter, $v_p$, is pivotal in quantifying the alteration in signal strength that Bob perceives when the MWS is strategically positioned at his reflection point within the NLOS link. Setting a threshold for $v_p$ at 0.5 implies that should Eve manage to disrupt more than half of Bob's signal, Bob would then become aware of a potential compromise in the channel's performance. The utility of NLOS transmission, particularly through signal reflection, is recognized as an essential strategy to address the inherent sparsity of the terahertz channel. Our MWS, specifically designed for enhancing the scattering link in NLOS communication, aim to disperse critical information towards potential eavesdroppers while still ensuring the integrity of the signal transmission to Bob.

However, it's important to note that the inherent roughness of reflective surfaces in our daily life may induce a reflection loss ranging from 0.5 dB to 40 dB, affecting the NLOS channel's efficacy across various indoor and outdoor settings [32-34]. This variability could lead to noticeable changes in the power level received by Bob, either as an increase or a decrease, post the surface's introduction. Nevertheless, such changes are typically within the permissible range of variation associated with standard surface reflection losses. Specifically, under conditions of rainfall, the attachment of raindrops to surfaces introduces additional variability, affecting the signal fluctuation range both before and after the precipitation. Fig. 3(b) illustrates the $v_p$ variation corresponding to typical surface reflection losses under both wet and dry conditions. Under dry scenarios, a reflection loss exceeding 9.9 dB results in a $v_p$ value below 0.5, whereas during rainfall, a reflection loss beyond 9.2 dB is necessary to achieve the same.

It's important to note that Fig. 3(a) restricts its focus to the scattering distribution at a fixed angle ($\theta_i = 45°$), aligning with surface rotation angle $\alpha = 0°$ as shown in Fig. 1(c). However, given that an eavesdropper is likely to exploit predetermined positions of Alice and Bob, manipulating the signal direction via surface rotation presents a strategic advantage. Fig. 4(a) showcases how $v_p$ varies as the surface undergoes rotation both clockwise and counterclockwise, with $\alpha > 0°$ indicating the direction as per Fig. 1(c). Remarkably, at 140 GHz, the MWS S2 showcases five distinct peaks, which indicates it can provide several opportunities for signal interception. Notably, at $\alpha = 0°$, the presence of rainfall reduces the $v_p$ value, initially above 0.5, to a lesser extent due to the MWS's reduced susceptibility to rain, thereby achieving a smaller ratio term in Eq. (2). Considering these observations, we can infer that Eve's potential to successfully eavesdrop is significantly enhanced when both the $c_s$ and $v_p$ values fall below 0.5 simultaneously, underscoring a methodical approach to intercepting communications effectively.

Successful eavesdropping at 140 GHz, irrespective of rainfall presence, points to surface rotation angles of $\alpha = 14°$ and $\alpha = 30°$ as optimal due to the smaller backscattering as discussed later. The angle $\alpha = 30°$ is particularly noteworthy due to its association with a relative larger $v_p$ value, indicating a reduced signal power at Bob's end, together with observable signal fluctuations. This finding corroborates the feasibility of eavesdropping at this and preceding angles. Following a subsequent rotation of the receiver by 30º to analyze the scattering distribution, the results depicted in Fig. 4(b) revealed that, aside from the direct reflection ($\theta_i = 75°$), multiple scattering peaks satisfy the eavesdropping threshold of $c_s < 0.5$. Although the signal remains decipherable, the variance among peaks under rainfall conditions is pronounced. With increasing incidence angles, the beam's footprint on the surface enlarges, heightening vulnerability to raindrop interference and potential signal loss beyond the surface's edge. This comparative analysis across various rotation angles, extended to other frequencies, further validates the setup's stability and resilience to environmental disturbances. Fig. 4(c) displays the $c_s$ values at 170 GHz under rainfall condition for surface rotation angles $\alpha = 25°$ and $\alpha = 38°$, respectively, identifying viable eavesdropping positions. The angle $\alpha = 38°$ corresponds to an incident angle of $\theta_i = 83°$, which introduces more serious signal degradation than that at $\alpha = 25°$. The strategic placement of the surface thus emerges as a critical factor in ensuring eavesdropping stability and underscores the surface's robustness against environmental adversities, affirming the feasibility of eavesdropping even amidst significant rainfall interference or signal attenuation. Furthermore, it can also be asserted that the MWS would maintain similar performance at higher frequencies, as the impact of raindrops on THz waves is consistently evident across the range of 100 GHz to 1 THz, with the exception of frequencies where atmospheric absorption peaks occur.

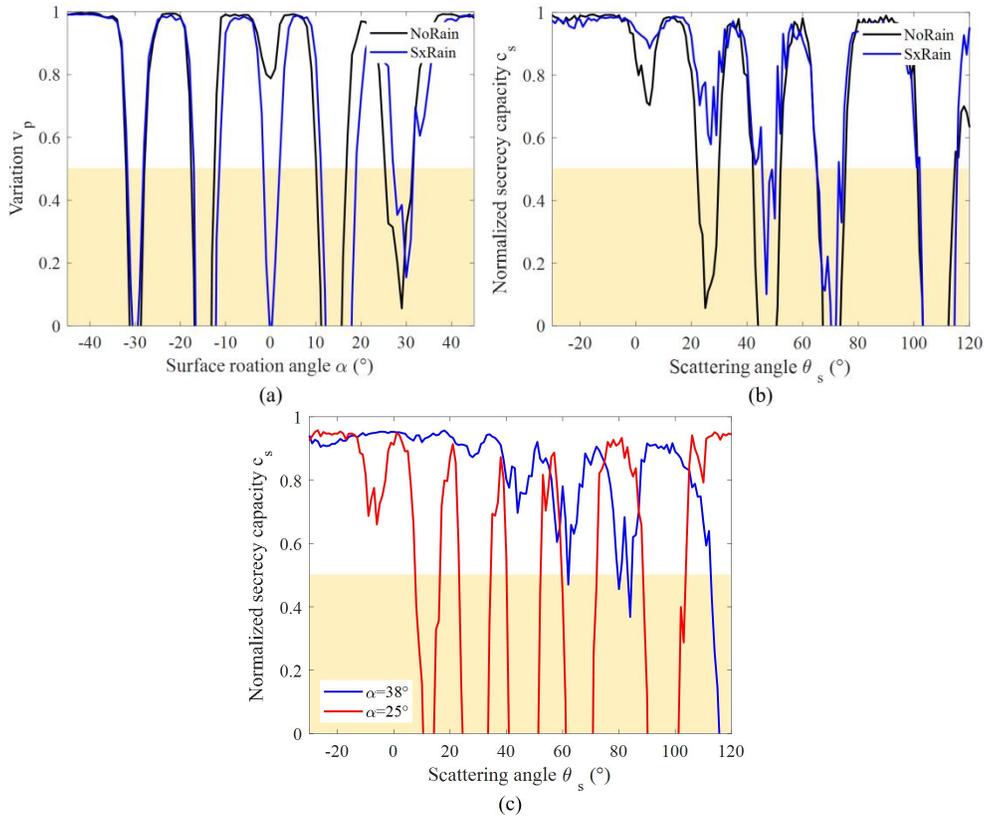

Fig. 4. (a) Measured *variation* parameter with respect to surface rotation in for the for an input frequency of 140 GHz with respect to surface rotation angle. (b) Measured angular positions for successful attacks when the surface rotation angle is set at 30º at 140 GHz. (c) Measured angular positions for successful attacks when the surface rotation angle is set at 25º and 38º at 170 GHz under rainfall condition. Yellow region means corresponding values below 0.5.

For an eavesdropping endeavor to be deemed successful, it must fulfill two critical criteria: the eavesdropper must intercept a strong enough signal without being detected by the intended receiver and must ensure that the signal transmitter does not pick up any backscattering warnings that could reveal the interference. The technique of backscattering measurement emerges as a vital tool for safety detection in this context [35]. This method hinges on the principle that when a signal bounces off a typically rough surface, a portion of it scatters back towards the source, Alice in our scenario. By effectively capturing and meticulously analyzing these backscattered signals, one can determine the presence of any anomalous signal interference. To this end, a specific *backscatter parameter*, denoted as $S_{180}$, has been established to gauge the success of this backscattering assessment [6].This parameter, expressed as

$$S_{180} = \left|1 - \frac{SNR_{Alice}^{nosurface}}{SNR_{Alice}^{surface}}\right| \qquad (3)$$

is instrumental in quantifying any changes in the signal received by Alice, setting a benchmark to identify deviations significant enough to raise an alert. A value of $S_{180}$ equal to 0 signifies no detectable change in the backscattered signal upon the introduction of an eavesdropper into the channel's reflection path. We designate $S_{180} = 0.5$ as the critical threshold for raising an alarm; values below this threshold suggest that the variations in the backscattered signal fall within expected, normal ranges of scattering.

In Fig. 5(a), the experimental setup outlines the transmitter and receiver aligned as closely as possible for measuring the backscattered signal. Due to equipment constraints, there is a slight angular discrepancy of about 15º between them. The experiment involves altering the surface's orientation at various incidence angles, aligning with the rotation angle $\alpha$ as depicted in Fig. 4(a), to measure backscattering. The findings reveal that most $S_{180}$ values exceed the 0.5 threshold when $\alpha < 0º$. This indicates a significant alteration in the backscattered signal, presumably detectable by Alice. However, in scenarios where $\alpha > 0º$, specifically for $\alpha \geq 14°$, the $S_{180}$ values drop below the 0.5 mark. And that's why we choose $\alpha = 30°$ for analysis in Fig. 4(b). This small $S_{180}$ value suggests that the backscattering changes remain undetectable to Alice, thereby marking the eavesdropping attempts as successful under these conditions. Thus, these observations substantiate that under specific angular adjustments, Eve's activities can proceed unnoticed, affirming the viability of the eavesdropping attack.

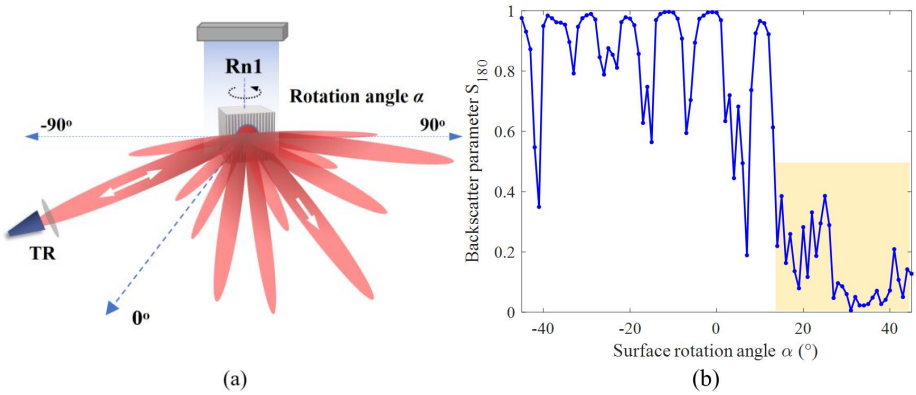

Fig. 5. (a) Schematic of the setup for the backscattering measurement under rainfall condition. (b) Values of the backscattering parameter $S_{180}$ measured for surface rotation under rainfall condition.

## 5. Conclusion

In this work, we explored the resilience of NLOS THz channels under the influence of rain, employing a weather emulation system alongside MWS. Positioned at key reflection points for NLOS transmission, the MWS encountered rainfall impacting both the direct transmission path and the surrounding area. Our analysis concentrated on how such conditions affect the channel's security, focusing on three critical aspects: *normalized secrecy capacity*, signal power *variation*, and *backscattering* phenomena. A notable discovery was the ability of the MWS design to modify the interaction between raindrops and itself, effectively reducing rain adhesion and lessening the deleterious effects of rainfall on the surface. The unique scattering properties of the MWS emerged as an important factor, ensuring that the fluctuations caused by rain-induced scattering did not overlap with the main signal peaks. This safeguarded the core transmission from significant distortion. Despite the challenging conditions posed by rainfall, our results demonstrate that eavesdropping attempts can still succeed, highlighting ongoing vulnerabilities in the physical layer security of outdoor NLOS communication links. These insights reveal the continuous need for vigilance and innovation in protecting against potential eavesdropping, emphasizing the importance of developing more sophisticated measures to secure terahertz communications against the elements and unauthorized interception.

**Funding**

This work was supported by the National Natural Science Foundation of China (62071046); the Science and Technology Innovation Program of Beijing Institute of Technology (2022CX01023); the Talent Support Program of Beijing Institute of Technology "Special Young Scholars" (3050011182153); and the Fundamental Research Funds for the Central Universities (2024CX06099).

**Author contributions**

**Peian Li**: Conceptualization, Methodology, Experiment, Software, Validation, Investigation, Writing. **Wenbo Liu**: Design, Methodology, Experiment, Investigation. **Da Li**: Experiment. **Mingxia Zhang**: Software. **Xiaopeng Wang**: Data provision. **Houjun Sun**: Investigation, Validation. **Jianjun Ma**: Conceptualization, Methodology, Experiment, Funding acquisition, Writing, Supervision.

## Disclosures

The authors declare no conflicts of interest.

## Data availability

Data underlying the results presented in this paper are not publicly available at this time but may be obtained from the authors upon reasonable request.